\documentclass[10pt,english,journal]{IEEEtran}
\usepackage[T1]{fontenc}
\usepackage[latin9]{inputenc}
\usepackage{geometry}
\usepackage{amsmath}
\usepackage{xpatch}
\usepackage{amssymb}
\usepackage{esint}
\usepackage{mathtools}
\usepackage{amsthm}
\usepackage{nicefrac}
\usepackage{bigints}
\usepackage{mathtools}
\usepackage{blkarray, bigstrut}
\usepackage{physics}
\usepackage{calligra}
\usepackage{graphicx}
\usepackage{epstopdf}
\usepackage{dsfont}
\usepackage{array,ragged2e}
\usepackage{enumitem}
\usepackage{etoolbox}
\usepackage{babel}
\usepackage[nopar]{lipsum}
\usepackage{psfrag}
\usepackage[ruled,norelsize]{algorithm2e}
\usepackage{algorithmic}
\usepackage{balance}
\usepackage{float}
\usepackage{subcaption}
\SetKw{KwBy}{by}

\makeatletter
\newcommand{\removelatexerror}{\let\@latex@error\@gobble}
\makeatother

\xpatchcmd{\proof}{\hskip\labelsep}{\hskip5\labelsep}{}{}  
\makeatletter
\xpatchcmd{\proof}{\@addpunct{.}}{\@addpunct{:}}{}{}
\makeatother

\renewcommand\[{\begin{equation}}
\renewcommand\]{\end{equation}} 
\usepackage{listings}
\usepackage{fancyvrb}
\usepackage{framed}

\usepackage{courier}
\usepackage[usenames,dvipsnames,table]{xcolor}

\definecolor{dkgreen}{rgb}{0,0.3,0}
\definecolor{gray}{rgb}{0.5,0.5,0.5}

\makeatletter
\newcommand*{\rom}[1]{\expandafter\@slowromancap\romannumeral #1@}
\makeatother

\usepackage{siunitx}
\usepackage{tabu}
\usepackage{booktabs}
\usepackage{multirow}

\usepackage{titlesec}
\usepackage{capt-of}
\usepackage{array}
\usepackage{arydshln}
\begin{document}
\title{A Mathematical Introduction to Deep Reinforcement Learning for 5G/6G Applications}
\author{Farhad Rezazadeh\\
{\normalsize{} Telecommunications Technological Center of Catalonia (CTTC), Barcelona, Spain}\\
{\normalsize{}Contact Emails: \texttt{frezazadeh@cttc.es}}\vspace{-0.1 cm}}
\maketitle
\thispagestyle{empty}
\begin{abstract}
Algorithmic innovation can unleash the potential of the beyond 5G (B5G)/6G communication systems. Artificial intelligence (AI)-driven zero-touch network slicing is envisaged as a promising cutting-edge technology to harness the full potential of heterogeneous 6G networks and enable the automation of demand-aware management and orchestration (MANO). The network slicing continues towards numerous slices with micro or macro services in 6G networks, and thereby, designing a robust, stable, and distributed learning mechanism is considered a necessity. In this regard, robust brain-inspired and dopamine-like learning methods, such as Actor-Critic approaches, can play a vital role. The tutorial begins with an introduction to network slicing, reinforcement learning (RL), and recent state-of-the-art (SoA) algorithms. Then, the paper elaborates on the combination of value-based and policy-based methods in the form of Actor-Critic techniques tailored to the needs of future wireless networks.

\end{abstract}


\section{Introduction}
\IEEEPARstart{N}{etwork} slicing creates an all-encompassing environment to underpin many network services by running fully or partly isolated logical networks on the same physical infrastructure. The zero-touch network is conceived as a next-generation of network management that leverages the principles of network function virtualization (NFV) and software-defined networking (SDN) to be the cornerstone for supporting fully automated operations and on-demand configuration without the need for fixed contractual agreements and manual intervention. To handle these radical changes, the zero-touch network and service management (ZSM) framework reference architecture \cite{ZSM} has been designed by ETSI. One of the essential building blocks in ZSM is intelligent decision-making elements. The quest for intelligent and optimal control in massive telecommunication environments has aroused intensive research on the applications of deep reinforcement learning (DRL) methods. The DRL can provide a promising technique to be incorporated in network slicing and solve the control and optimization issues.

DRL combines RL and deep neural network (DNN) to extract knowledge based on gained experience by interacting with the environment (network). To solve slice issues such as admission control, congestion control, energy efficiency, resource management, service creation, dynamic network configuration, anomaly and fault detection, security, and reliability, the agent of network slicing should update the Q-function for optimal actions. Deep Q-learning benefits from a neural network in approximating the Q-value function. The state can be considered input, and the Q-value of all possible actions is generated as output. An agent (neural network) in DRL generates the dataset on the fly and continuously interacts with a network well-suited to problems with numerous possible states with high-dimension. In complex wireless networks such as network slicing, the state and action spaces are large, and thereby, RL becomes unstable or fails to converge. We use a function approximation like a neural network to calculate the Q-values. In essence, DRL has great potential for handling the extensive uncertainty and dynamic nature of network slicing with a deep neural network. However, the DRL methods such as deep Q-network (DQN) enable to solve the problems of networks with high-dimensional observation spaces (states), they are unable to handle high-dimensional and continuous action spaces (e.g., resource allocation, traffic optimization, and energy control.) where the number of actions can increase exponentially in massive slices environment. Here, we can combine DQN and policy gradients in the Actor-Critic approach with insights from the recent success of continuous action and state spaces in network slicing \cite{ref2}.

Figure 1 shows the extended version of the functional model for decision-making in the circuit linking the cerebral cortex and the basal ganglia in the human brain \cite{ref3}. 
\begin{figure}[h!]
\centering
\includegraphics[scale=0.45]{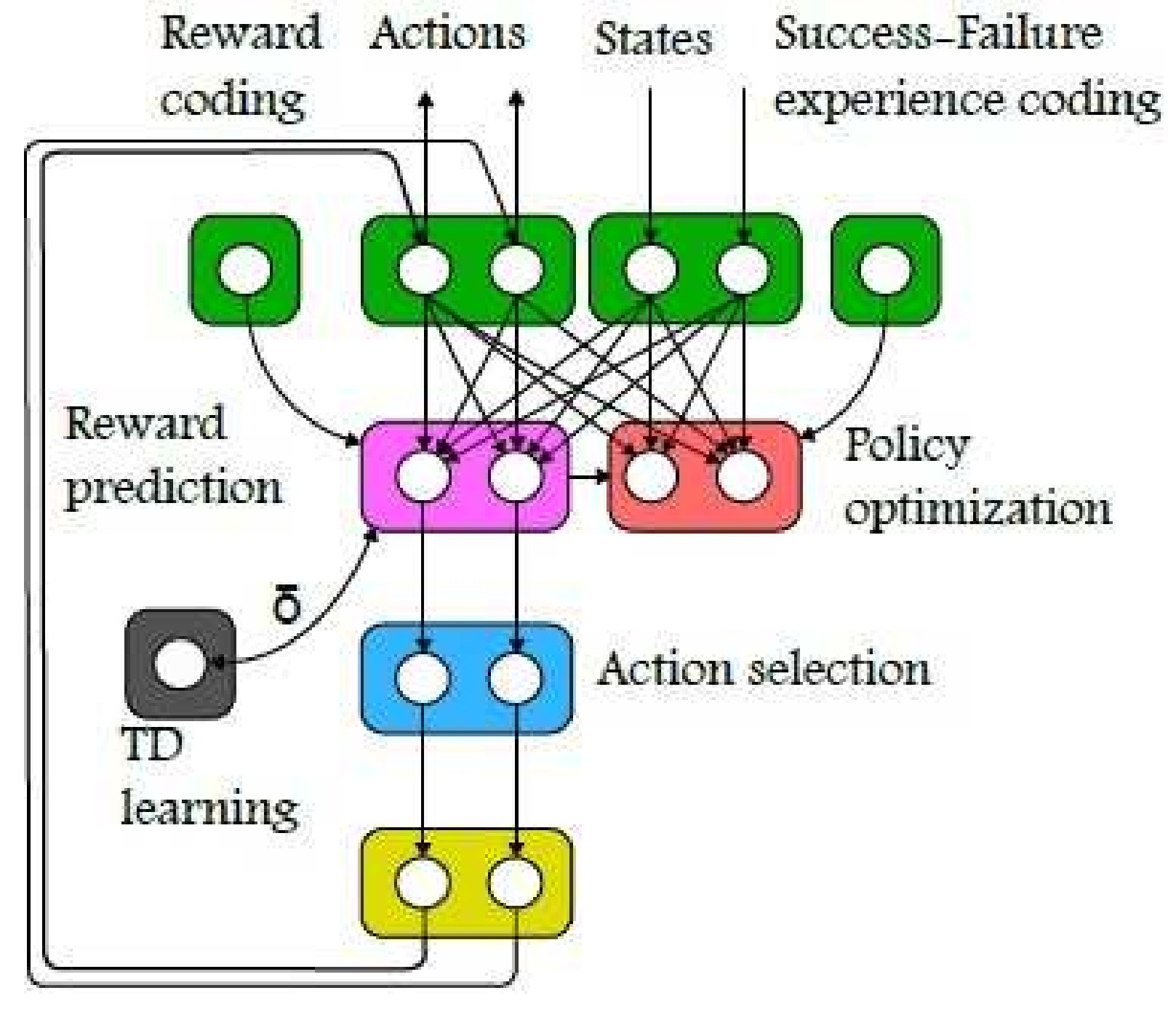}
\caption{The functional model for RL in the human brain.}
\end{figure}
In this context, Barto and Sutton \cite{ref4} have proposed an RL method called Actor-Critic. Figure 2 demonstrates the architecture of the Actor-Critic method. The idea of calculating value function by a second neural network as the baseline is called the Actor-Critic method, where the policy network $(\pi)$ is the actor and value network $(V)$ is the critic. Indeed, Actor-Critic is inspired by dopamine-like learning in the human brain to signal anticipation of future rewards to reinforce specific actions.
\begin{figure}[h!]
\centering
\includegraphics[scale=0.45]{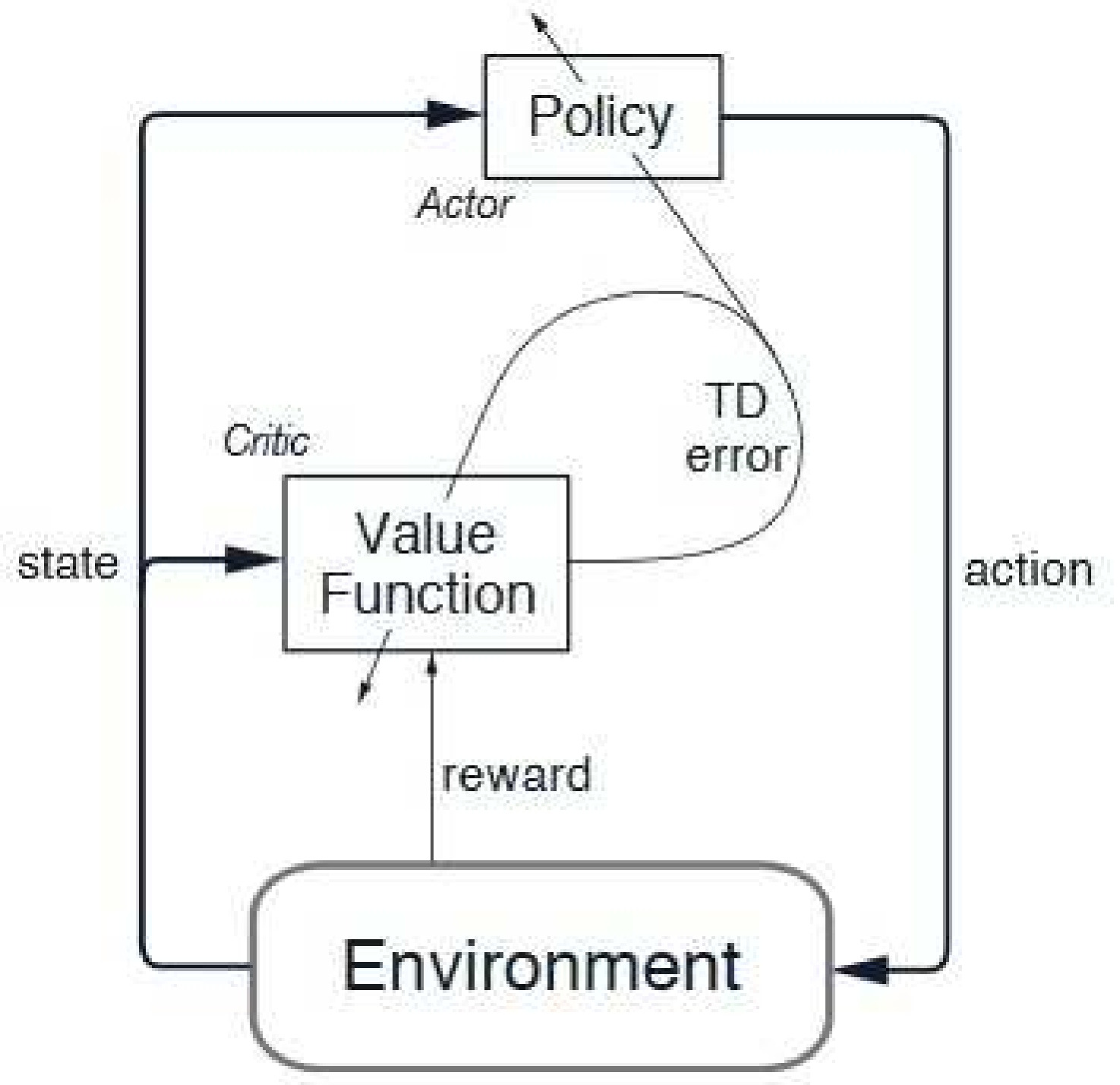}
\caption{The architecture of Actor-Critic.}
\end{figure}

In \cite{ref5}, the authors have investigated a demand-aware inter-slice resource management solution based on advantage Actor-Critic (A2C) as a DRL algorithm to track the user mobility and improve the system utility aiming to make appropriate resource allocation decisions. In \cite{ref6}, the authors have proposed a scheme to allocate network resources effectively. The authors have integrated the alternating direction method of multipliers (ADMM) and DRL, where they exploit the deep deterministic policy gradient (DDPG) as an Actor-Critic technique to learn the optimal policy. Pujol Roig \emph{et al.} have proposed an Actor-Critic, called parameterized action twin (PAT) deterministic policy gradient algorithm where automated MANO allows a central unit (CU) to learn to re-configure resources autonomously \cite{ref7}. Liu \emph{et al.} have proposed a DRL-based method called DeepSlicing, where they decompose the network slicing problem into a master problem and several slave problems wherein Actor-Critic agents learn the optimal resource allocation policy \cite{ref8}. Li \emph{et al.} have proposed a DDPG and Actor-Critic-based solution to enhance energy efficiency and obtain the optimal power control scheme \cite{ref9}. Correspondingly, \cite{ref10} has proposed an Actor-Critic method, called the twin-delayed double-Q soft Actor-Critic (TDSAC) method, to minimize energy consumption and virtual network function (VNF) instantiation cost in network slicing. In \cite{ref11}, the authors have proposed an Actor-Critic-based network slicing method called TD3 \cite{ref12}. They have introduced a multi-objective approach to make the CU learn how to re-configure computing resources autonomously while minimizing latency, energy consumption, and VNF instantiation cost for each slice. From this SoA overview, more efforts should be deployed on RL methods that support high-dimensional state space while avoiding the curse of dimensionality, gradient explosion, and catastrophic forgetting problems. The role of Actor-Critic and continuous action approaches should also be investigated as more stable mechanisms in real environments. Besides, the design of robust Actor-Critic algorithms is considered necessary in 6G.

\section{Distributed Actor-Critic Learning}
\subsection{Reinforcement learning}
RL technique is evaluative feedback-based learning and a branch of machine learning (ML) to optimize the accumulated long-term reward (average return). It automates complex sequential decision-making tasks of an agent without a priori knowledge of the system through interaction with the environment. To enable the agent to take the right actions, it should spend time in a certain part of the environment. Generally, all the sequences of actions, rewards, and observation time up to time step $t$ is called the history
\begin{equation}
       H_t = [O_1, A_1, R_1, ..., O_t, A_t, R_t].
\end{equation}

Markov decision process (MDP) 's mathematical context can be considered a formal framework to formalize diverse RL methods for learning optimum decision-making policies in a fully or partially observable environment. One of the important issues is the intelligent agent can dynamically make decisions based on actions, states, and rewards where the states refer to possible network configurations, and reward (or penalty) stands for feedback signal from the network (environment) that implies the agent's performance. The agent observes the network state at each time step $t$ and acts on that network to transit from one state to another. Typically, an MDP is defined by a 5-tuple $(S, A, P, \gamma, R)$ where $S$ is a set of state (state space), $A$ refers to a set of action (action space), $P$ denotes the transition probability from current state $s$ to the next state $s^\prime$ where they govern rules of state transitions and define our dynamics. The $R$ notation stands for the reward function. The agent obtains an immediate reward by taking action $a_t$ in state $s_t$. Unlike finite or episodic tasks, continuing tasks can have infinite total rewards. The $\gamma$ defines a reward discounting hyperparameter that results in significant deviations in the agent's performance where the future agent values rewards will turn to less and less. Indeed, it is a real-valued discount factor weighting to determine the importance of future rewards as a short-sighted or myopic agent $(\gamma=0)$, i.e., agent aims to maximize its current/immediate rewards, or far-sighted agent $(\gamma=1)$ that strives to accumulate long-term higher rewards. The $\gamma$ is often chosen in $[0.95, 0.99]$. The total discounted rewards from time step $t$ is given by
\begin{equation}
{G}_t = R_{t+1}+\gamma R_{t+2}+\gamma^2 R_{t+3}+ ...  \sum_{n=0}^{\infty}\left(\gamma^{n} R_{t+n+1}\right).
\end{equation}
We can derive a useful recursive relation between rewards and subsequent time step, $G_t = R_{t+1} + \gamma G_{t+1}$ that can attenuate computational complexity and memory requirements. However, there is one more piece we need to complete the MDP. The key term of MDP is decision. Indeed, the way we make decisions for what actions to do in what states is called a policy, which denotes with the symbol $\pi$ that it is usually a probabilistic function as a mapping of states to actions.

Technically, the policy is not part of the MDP itself but part of the solution along with the value function. Unlike supervised and unsupervised learning, the agent in RL banks on reward signals to evaluate the effectiveness of actions and enhance the improvement of the policy. The excellence of each state or state-action pair can be determined by how large the future reward of the agent is. The state value function for the policy $\pi$ is an explicit measure of how good the state is or how much reward to expect
\begin{equation}
{V}_{\pi}(s)= \mathbb{E}_{\pi}\left[\sum_{n=0}^{\infty}\left(\gamma^{n} R_{t+n+1} | S_t = s \right)\right],
\end{equation}
and expectation value of the reward at time $t$ is defined as the action value function
\begin{equation}
{Q}_{\pi}(s,a) = \mathbb{E}_{\pi}\left[\sum_{n=0}^{\infty}\left(\gamma^{n} R_{t+n+1} | S_t = s, A_t = a \right)\right].
\end{equation}
According to total expectation in reverse and nested expected value, we can define the Bellman equation by the assumption of MDP
\begin{equation}
{V}_{\pi}(s)= \sum_{a}\pi(a|s)\sum_{s^\prime}\sum_{r}p(s^\prime, r | s,a)[r + \gamma{V}_{\pi}(s^\prime)].
\end{equation}
The above recursive relationship between each value function and the successor state's value function is considered the heart of RL to solve stochastic and non-deterministic search problems where agents encounter random events. In deterministic policy, there is just one action with $\pi(a|s) = 1$, and the rest receive $0$. Some algorithms specifically apply to only the state-value function or only the action-value function. Let suppose we have two policies $\pi_1$ and $\pi_2$ and $\pi_1$ is better than $\pi_2$, if expected return of $\pi_1$, $V_{\pi_1}(s)$ greater than or equal to the expected return of $\pi_2$, $V_{\pi_2}(s)$ for all states
\begin{equation}
\forall\quad s \in S,\quad if\quad V_{\pi_1}(s) \geq V_{\pi_2}(s)\quad \rightarrow \quad\pi_1 \geq \pi_2
\end{equation}

The optimal policy in RL is the best one for which no greater value function exists. The optimal value functions are optimal, but optimal policies are not necessarily unique. Due to designing the RL program, there are two different problems we should try to solve. The finding $V(s)$ given a policy is called the prediction problem, and finding the optimal policy for a given environment refers to control problems. The prediction problem is easier than the control problem, and the job is to find the value function based on the algorithm.

In the parlance of RL, the agent engages in exploration and exploitation, where exploitation or greed refers to taking the best-known action while sub-optimal action is called exploration (i.e., the way of exploring the environment). This process of exploration and exploitation enables the agent to refine its model and gradually find the true estimate of the value function. The control algorithms with a model-based (planning) approach constitute a set of distributed computations in the network to build a model of how the environment works, which refers to agents' decision-making in the control theory domain. The agents don't need to learn network dynamics to adopt this scheme, but it is necessary to assume some network parameters and configurations (e.g., channel information). In contrast, black-box network optimization/control works without a priori network model, namely model-free methods where it relies on a trial and error approach to learn the state transition probabilities to solve the Bellman equation and try to figure out a policy of how to behave optimally. At the same time, this trade-off of time aims to ease computational complexity and yield controller convergence. The tendency towards fulfilling fully automated operations and limited computational resources in wireless networks has aroused intensive research interest in working on model-free problems.

Q-learning is a model-free approach in the class of temporal difference (TD) learning algorithms where it has an online nature and updates the agents' estimate of value function at each time step. To update the value function concerning TD, we have
\begin{equation}
{V}(s_t)= {V}(s_t) + \alpha [R_{t+1}+\gamma V(s_{t+1}) - V(s_t)]
\end{equation}
where  $\alpha$ denotes an exponentially smooth mean approach to deal with non-stationary distributions instead of calculating the sample mean inefficiently, and $R_{t+1}+\gamma V(s_{t+1})$ is the target of the TD method and $V(S_t)$ refers to old estimate. Moreover, the quantity in the brackets $R_{t+1}+\gamma V(s_{t+1}) - V(S_t)$ is called TD error. Therefore, we have squared error $E = [Target-Prediction]^2$ when pursuing a gradient descent approach, and the aim is for the prediction value to be closer to the target value. This approach that uses one estimate to update another estimate is called bootstrapping. The updated Q-value (quantifiable value for more lucrative action) in Q-learning as off-policy method is given by
\begin{multline}
Q(s_t,a_t)=
Q(s_t,a_t) +\alpha [R_{t+1}+\gamma \underset{a}{\max}Q(s_{t+1}, a_{max})\\
- Q(s_t,a_t)].
\end{multline}
Q-learning alternates between policy evaluation and, thereby, policy improvement, where an action greedily concerns the current value estimate. Unlike on-policy methods, in which one policy generates actions and updates the value function, Q-learning is an off-policy method that uses one policy to generate actions and another policy to update the value function. For example, Q-learning uses $\epsilon-greedy$ policy to choose an action while it updates the action-value function according to purely greedy action.

\subsection{Deep Q-Learning (DQN)}
The Q-learning works in a very simplistic environment, while in practice, communication network problems have complicated system models with large and continuous state spaces that can have infinite states. Therefore, the tabular representation of the previous section's action-value function leads to computationally complex and training time limitation problems. Unlike the tabular and non-parametric approach, RL is assisted with DNN in DRL to surmount the curse of dimensionality concerning inordinate large state spaces. Indeed, it is possible to model $Q(s, a)$ based on simple linear regression, but another new approach is the DNN that benefits from crafting inductive biases to overcome the curse of dimensionality effectively and thereby enables RL to scale the decision-making problems for intractable high-dimensional state and action spaces through approximate the Q-value function. The DQN takes the state as input and returns approximated Q-functions of all actions under the input state. In this regard, finding a convincing solution for the instability of function approximation techniques in RL-based agents is necessary.

 In this regard, we parameterize the value function with linear regression or a neural network. Let define $V(s) = \theta^T s$ where $s$ is a feature vector that represents the state, and in linear regression for calculating the value function prediction, $\theta$ refers to the weights. According to the previous section, we apply gradient descent using squared error between target and prediction. Instead of updating $V(s)$ directly, the goal is to update the parameters of $V$ that, according to the chain rule and multiply to the gradient of $V(s)$ we have 
\begin{equation}
\theta \leftarrow \theta + \alpha [r+\gamma V(s^\prime) - V(s)]\frac{\partial V(s)}{\partial\theta}
\end{equation}
and in linear regression, we have a simple update rule 
\begin{equation}
\theta \leftarrow \theta + \alpha [r+\gamma V(s^\prime) - V(s)]s
\end{equation}
with the same logic for $Q-updating$, we just consider $s$ as input, and for infinite action, the number of outputs is equal to the number of possible actions
\begin{equation}
\theta \leftarrow \theta + \alpha [r+\gamma \underset{a}{\max}Q(s^\prime, a_{max}) - Q(s,a)]\frac{\partial Q(s,a)}{\partial\theta}.
\end{equation}

While the DQN approach effectively solves communication network problems~\cite{Specialization_TVT} with large and continuous state spaces, it cannot support infinite or continuous action spaces. However, DRL can solve issues with gradient-free approaches, making it suitable for such scenarios.

\subsection{Policy Gradient}
Another approach to solving the control problem is called the policy gradient method, where we also parameterize the policy. Q-learning is a value-based method where the policy is to take the action that gives the best value 
\begin{equation}
a^* = \underset{a}{\arg\max}Q(s,a).
\end{equation}

In this case, the return is a probability distribution and determines the best action for a given state $\pi(a | s)$. Indeed, we sample from $\pi(a | s)$ because the policy is probabilistic, and there is no need to pursue strategies such as $\epsilon-greedy$. It creates more flexibility and chance for exploration with sampling from policy distribution and more control of the amount of randomness. In deep Q-learning, the neural network has one output for each action, thereby supporting infinite or discrete action spaces. In contrast, the output of policy networks is a probability distribution. Unlike the value-based method, in the policy-based method, instead of considering objective function directly, the main objective is the integral of gradient called policy gradient. The aim is to maximize the sum of total rewards, but there is no functional dependence on the policy parameters $\theta$, and thereby, there is no formula where reward depends on neural network weights. Let us define $\psi =\{(s_1, a_1), (s_2, a_2), ..., (s_T, a_T)\}$ as sequence of state-action pairs in an episode and define $G$ as sum of rewards where the aim is maximize the return $G(\psi) = \sum_{t=1}^TR(s_t, a_t)$. We define objective concerning $G$
\begin{equation}
J(\theta) = E [G(\psi), \pi_{\theta}].
\end{equation}
In fact, objective $J$ is a function of neural network weights $\theta$, and $\pi_{\theta}$ refers to the neural network output as policy distribution. The weights in expected value are probabilities, so we can rewrite expected value according to probability
\begin{equation}
J(\theta) =\sum_{\psi}P(\psi,\theta)G(\psi).
\end{equation}
Notice that just $P$ depends on $\theta$. Due to the derivative of $J$ we can move the gradient symbol inside the sum where $G$ is a constant concerning $\theta$
\begin{equation}
\nabla_{\theta}J(\theta) =\sum_{\psi}\nabla_{\theta}P(\psi,\theta)G(\psi).
\end{equation}
We can update the above equation according to 
\begin{equation}
\nabla_{\theta}J(\theta) =\sum_{\psi}\frac{P(\psi,\theta)}{P(\psi,\theta)}\nabla_{\theta}P(\psi,\theta)G(\psi).
\end{equation}
With respect to apply chain rule $\nabla\log f = \frac{\nabla f}{f}$ we have 
\begin{equation}
\nabla_{\theta}J(\theta) =\sum_{\psi}{P(\psi,\theta)}\nabla_{\theta} \log P(\psi,\theta)G(\psi).
\end{equation}
Therefore, according to expected values in reverse, we can define the gradient of the objective in terms of $\theta$ and neural network parameters
\begin{equation}
\nabla_{\theta}J(\theta) = E [\nabla_{\theta} \log P(\psi,\theta)G(\psi)],
\end{equation}
thereby, if $G$ is positive, then reinforce and lead actions to it, whereas if $G$ is negative, then avoid the tendency to take action by making it less probable. The problem is we cannot calculate the sum of infinite $s$ of action-state pairs or trajectories, so we can pursue a sampling approach where we can calculate the sample mean of some collected data that called rollouts
\begin{equation}
\nabla_{\theta}J(\theta) \approx \frac{1}{N}\sum_{i = 1}^N\nabla_{\theta} \log P(\psi_i,\theta)G(\psi_i).
\end{equation}
We should clarify how $P$ is dependent on $\pi$ and $\theta$. In these terms, we have an MDP problem where it is a Markov chain consisting of dynamics model, and also a policy that we can control it, and thereby, the probability is a simple Markov chain, $
P(\psi,\theta) = \Pi_{t = 1}^T p(s_{t+1} | s_t, a_t)\pi_{\theta}(a_t | s_t)
$. According to Eqn. 18 and calculate the gradient of the log we have, $
\nabla_{\theta} \log P(\psi,\theta) = \sum_{t = 1}^T \nabla_{\theta} \log\pi_{\theta}(a_t | s_t).
$ Notice that $\sum_{t = 1}^T \nabla_{\theta} \log p(s_{t+1} | s_t, a_t) = 0$ because it dose not depend on parameter $\theta$. Let us update Eqn. 18 
\begin{equation}
\nabla_{\theta}J(\theta) = E [\sum_{t = 1}^T \nabla_{\theta} \log\pi_{\theta}(a_t | s_t)G(\psi)].
\end{equation}
Due to ignoring full trajectories concerning $\psi$, we can define $G(\psi)$ as the sum of past and future rewards. We know the rewards that have performed before $a_t$ can not have an impact on policy and gradient descent because $a_t$ does not affect previous rewards, so with regards to discount future rewards, we have
\begin{equation}
\nabla_{\theta}J(\theta) = E [\sum_{t = 1}^T \nabla_{\theta} \log\pi_{\theta}(a_t | s_t)\sum_{j = t+1}^T \gamma^{j-t-1}R(s_j, a_j)].
\end{equation}
As mentioned before, it is time-consuming or maybe impossible to calculate future rewards in a complete episode when we have infinite horizon MDP so that we can replace it with action value $Q$ 
\begin{equation}
\nabla_{\theta}J(\theta) = E [\sum_{t = 1}^T \nabla_{\theta} \log\pi_{\theta}(a_t | s_t)Q(s_t, a_t)].
\end{equation}

Moreover, we can replace the future rewards with an N-step estimate approach where we sample the first $N$ rewards and estimate the rest through value function $V(s)$
\begin{equation}
R_t^{(N)} = R_{t+1}+ \gamma R_{t+2}+ ... + \gamma^{N-1}R_{t+N}+ \gamma^N V(S_{t+N}).
\end{equation}
In terms of baseline modification, let us define $A(s, a) = Q(s, a) - v(s)$ as an advantage value, which can be rewritten according to the Bellman optimality equation as $A(s, a) = R+ \gamma V(s^\prime) - V(s)$. Therefore, we use advantage instead of using future rewards 
\begin{equation}
\nabla_{\theta}J(\theta) = E [\sum_{t = 1}^T \nabla_{\theta} \log\pi_{\theta}(a_t | s_t)A(s_t, a_t)].
\end{equation}

We pursue this approach to improve policy gradients while reducing variance with a baseline to increase stability because we don't care about the absolute reward. To calculate $V(s)$, we use another neural network.

\subsection{Distributed Learning in Actor-Critic}
The Actor-Critic method is a combination of DQN and policy optimization. Usually, when we use the neural network in classic RL methods, it is not robust enough because of too much variance in training data. Therefore, recovering the procedure concerning bad data or wrong direction is difficult. A strategy to solve this problem and reduce the variance is to use multiple parallel rollouts. In fact, instead of looking at a sample in stochastic gradient descent, we can pursue a batch gradient descent where we consider multiple samples simultaneously to stabilize the learning. As we mentioned before, in the Actor-Critic method, the actor is a neural network for parameterizing the policy $\pi(a | s, \theta_p)$ as output, and the critic is another neural network to parameterize the value function  $V(s, \theta_v)$ as output where $\theta$ denotes the weights and the $Advantage$ is difference between the actual return and the value, $Advantage = G - V(s)$. To calculate policy loss with a minimized approach, we use backward from policy gradient or integral of the gradient  
\begin{equation}
L_p = - (G-V(s))\log\pi(a| s, \theta_p),
\end{equation}
and for value network loss, we use squared error
\begin{equation}
L_v = (G-V(s, \theta_v))^2,
\end{equation}
where the target can be a TD return or an N-step return. We use N-step returns when collecting multiple samples before doing the update. For updating $\theta_v$, we make gradient descent on squared error,
\begin{equation}
    \theta_v \leftarrow  \theta_v - learning\_rate * \nabla_{\theta_v}L_v.
\end{equation}

\begin{table*}
\caption{Comparison of architecture and hyperparameter tuning between algorithms.}
\scriptsize
\centering
\begin{tabular}{@{}lccccccccc@{}}\toprule
\textbf{Architecture} & \textbf{DDPG} &  \textbf{SAC} & \textbf{TD3}\\ \midrule
\textbf{Method} & Actor-Critic  & Actor-Critic  & Actor-Critic\\ \hdashline
\textbf{Model Type} & Multilayer perceptron  & Multilayer perceptron  & Multilayer perceptron \\ \hdashline

\textbf{Policy Type} & Deterministic &  Stochastic &  Deterministic 
\\ \hdashline
\textbf{Policy Evaluation} & TD learning &  Clipped double Q-learning & Clipped double Q-learning \\ \hdashline
\textbf{No. of DNNs}& 4  & 5  & 6   \\ \hdashline
\textbf{No. of Policy DNNs}& 1  & 1  & 1  \\ \hdashline
\textbf{No. of Value DNNs}& 1  & 2  & 2   \\ \hdashline
\textbf{No. of Target DNNs}& 2  & 2  & 3 \\ \hdashline
\textbf{No. of hidden layers}& 2  & 2  & 2   \\ \hdashline
\textbf{No. of hidden units/layer}& 200  & 256 &  400/300\\ \hdashline
\textbf{No. of Time Steps}& based on problem &  based on problem &  based on problem\\ \hdashline
\textbf{Batch Size}& 64  & 256  & 100   \\ \hdashline
\textbf{Optimizer}& ADAM &  ADAM &  ADAM  \\ \hdashline
\textbf{ADAM Parameters ($\beta_1, \beta_2$)}& (0.9, 0.999) &  (0.9, 0.999)  &  (0.9, 0.999)   \\ \hdashline

\textbf{Nonlinearity}& ReLU  & ReLU & ReLU\\ \hdashline
\textbf{Target Smoothing $(\tau)$}& 0.001 &  0.005  & 0.005  \\ \hdashline
\textbf{Exploration Noise}& $\theta,\sigma=0.15,0.2$  & None &  $\mathcal{N}(0,0.1)$ \\ \hdashline

\textbf{Update Interval}& None & None  & 2\\ \hdashline
\textbf{Policy Smoothing}& None  & None  & $\epsilon \sim clip(\mathcal{N}(0,0.2), -0.5, 0.5)$ \\ \hdashline
\textbf{Expected Entropy$(\mathcal{H})$}& None & -dim(Action) & None\\ \hdashline
\textbf{Actor Learning Rate}& 0.0001 &  0.0001  & 0.001  \\ \hdashline
\textbf{Critic Learning Rate}& 0.001 & 0.0001 &  0.001 \\ \hdashline
\textbf{Reward Scaling}& 1.0  & 0.2  & 1.0 \\ \hdashline

\textbf{Discount Factor}& 0.99  & 0.99  & 0.99  \\ \hdashline
\textbf{Replay Buffer Type}& Simple &  Simple &  Simple  \\ \hdashline
\textbf{Replay Buffer Size}& based on problem &  based on problem &  based on problem \\ \hdashline
\textbf{Max Episode Length}& based on problem &  based on problem &  based on problem  \\
\hdashline
\textbf{Seed}& System time &  System time &  System time \\
\hdashline

\bottomrule
\end{tabular}

\end{table*}

The A3C method, a parallelization approach, significantly reduces computation time. In this method, multiple workers (agents) interact asynchronously with their own environments to collect training data. Each worker has a copy of the policy and value networks, updated based on the current weights. The worker then sends its gradient to the master network, which updates its parameters and sends them back to the worker. This parallelization approach not only saves time but also enhances stability.

The worker networks usually have a recent copy of the master network weights. The problem with this method is some agents need to be updated. The asynchrony approach can lead some agents to play with an older version of the parameters. In contrast, the global network can get many updates in parallel ways, saving time and achieving more stability because the workers play different episodes with the same parameters and get more diverse data. Another method is Advantage Actor-Critic (A2C); the key difference from A2C is the Asynchronous part. In synchronous mode, we need just one copy of the neural network, whereas in A3C, we have multiple copies.

The modern Actor-Critic approaches are classified according to DDPG \cite{ref13}, TD3 \cite{ref12}, and SAC \cite{ref14} methods. Table \rom{1} compares these algorithms regarding architecture and hyperparameter tuning. The TD3 is a developed version of DDPG. The policy type in TD3 and DDPG is deterministic \cite{ref15}. In \cite{ref16}-\cite{ref17} have mentioned that deterministic approach and model-free DRL can have an extreme brittleness and hyperparameter sensitivity nature because hyperparameters should be tuned for a specific domain, and thereby, it can be problematic to deploy a one-size-fits-all approach. We use a stochastic policy gradient approach with a maximum entropy RL framework to mitigate the high sample complexity and the necessity of meticulous hyperparameter tuning in the SAC method. The entropy concept is borrowed from physics, where it has to handle the amount of disorder in the system.

\section{Conclusion}
We have deliberated on algorithmic innovation, an AI-driven approach, and continuous model-free DRL methods to solve control problems in telecommunication and network slicing. Despite the successes of Actor-Critic methods, many issues need to be addressed before these techniques can be applied to a wide range of complex real problems in B5G/6G. The issues can include reliability and time efficiency for dynamic network access, admission control, resource allocation, data rate control, wireless caching, network security, data offloading, traffic routing, and other control problems.

\end{document}